\documentclass[12pt]{iopart}

\usepackage{graphicx}% Include figure files
\usepackage{dcolumn}% Align table columns on decimal point
\usepackage{bm}% bold math
\usepackage{tabularx}
\usepackage{lineno}
\usepackage{float}
\newcommand{\bef}{\begin{figure}}
\newcommand{\eef}{\end{figure}}

\newcommand{\be}{\begin{equation}}
\newcommand{\ee}{\end{equation}}
\newcommand{\bea}{\begin{eqnarray}}
\newcommand{\eea}{\end{eqnarray}}
\begin{document}

\title[]{Insight from elliptic flow of open charm mesons using quark coalescence model at RHIC and LHC energies}

\author{Roli Esha, Md. Nasim and Huan Zhong Huang}

\address{University of California, Los Angeles, USA}
\ead{roliesha@physics.ucla.edu, mdnasim@physics.ucla.edu and huang@physics.ucla.edu}
\vspace{10pt}
%\begin{indented}
%\item[]December 2016
%\end{indented}
%\linenumbers

\begin{abstract}
A study of elliptic flow of open charm mesons, $D^0$ and $D_S ^\pm$
using quark coalescence as a mechanism of hadronization of heavy
quarks implemented in conjunction with A Multi Phase Transport (AMPT)
model has been presented. We have studied the transverse momentum
dependence of the elliptic flow parameter at mid-rapidity ($|y|$ $<$
1.0) for Au+Au collisions at $\sqrt{s_{NN}} = 200$ GeV (RHIC) and Pb+Pb collisions at $\sqrt{s_{NN}} = 2.76$ TeV (LHC) for different values of partonic interaction cross-section and QCD coupling constant. We have compared our calculations with the experimentally measured data at the LHC energy. We have also studied the effect of shear viscosity on elliptic flow of open charm mesons within the transport model approach. Our study indicates that the elliptic flow of open charmed mesons is more sensitive to viscous properties of QGP medium as compared to light  charged hadrons. 
\end{abstract}

\vspace{2pc}
\noindent{\it Keywords}: open charm meson, elliptic flow,  coalescence
%\submitto{\JPG}
%\maketitle
%\linenumbers

\section{Introduction}
The initial spatial anisotropy of the collision geometry and the interaction among particles produced in high energy heavy ion collisions lead to anisotropy in the momentum of the produced particles~\cite{v2_1,v2_2,v2_3,v2_4}. This momentum anisotropy is manifested in the azimuthal distribution of particles with respect to the reaction plane. The azimuthal distribution of particles relative to reaction plane can be written as a Fourier series
\be
\frac{dN}{d \phi} = \frac{1}{2 \pi} \left( 1 + \sum_{n = 1} ^\infty 2 v_n \cos \left[n(\phi-\Psi)\right] \right),
\label{1}
\ee
where $\Psi$ is the reaction plane angle and $\phi$ is azimuthal emission angle of produced particles.

The second harmonic of this Fourier expansion is called elliptic flow. The elliptic flow parameter, $v_{2}$, is a good variable for studying the medium formed in the high energy heavy ion collisions~\cite{hydro,hydro1,hydro2,hydro3,hydro4,early_v2}. The conversion of geometrical eccentricity to elliptic flow depends on the transport properties and the equation of state of the medium. The magnitude of $v_2$ is estimated by ~\cite{method,method2} 
\begin{equation}
v_{2}=\langle\cos(2(\phi-\Psi))\rangle,
\label{2}
\end{equation}
where the $\langle\rangle$ denotes the average over all particles in all events. Equation~\ref{1} and~\ref{2} are applicable only in the idealistic case when event-by-event flow fluctuations can be neglected. In reality, one should use different event planes ($\Psi_{n}$) for different harmonics. Event plane is an estimation of the reaction plane by using the anisotropic flow itself ~\cite{method} .\\
Understanding the physics of strong interaction requires the understanding of interactions among its building blocks -- quarks and gluons. Heavy quarks are considered as an important probe to understand the properties of quark-gluon plasma (QGP) created in relativistic heavy ion collisions~\cite{heavy_1,heavy_2,heavy_3,heavy_4,heavy_5,heavy_6,heavy_7,ds_prl}. These are produced on a short time scale ($\sim$0.08 fm/c for $c\bar{c}$ production) in hard partonic scatterings during the early stages of the nucleus--nucleus collision. The probability of thermal production of heavy quark pairs in the high temperature phase of the QGP is expected to be small in existing accelerator experiments. Therefore, the total number of charm quarks is frozen very early in the history of the collision. The measurement of various properties of charmed meson ($D^{0}$, $D^{\pm}$, $D^{\pm}_{S}$ etc.) like elliptic flow is expected to reflect information from very early stage of the collision. 

Among all the open charm mesons, the charm-strange mesons, $D_{S}^{\pm}$, is identified as a particularly sensitive probe for the hot nuclear medium because of its unique valence quark composition. The production of $D^{\pm}_{S}$  can be influenced by the charm quark recombination with strange partons through enhanced thermal production of strange quarks in the deconfined matter~\cite{ds_prl,ds_old}. Quite recently, there have been several efforts both at RHIC and LHC to understand the properties of open charm mesons produced in high energy heavy ion collisions~\cite{star_d0,star_ds,alice_d0,alice_ds}. 

In this paper, we have systematically studied $v_{2}$ of $D^{0}$ and $D^{\pm}_{S}$  at $\sqrt{s_{NN}}$ = 200 GeV and 2.76 TeV using A Multi Phase Transport (AMPT) model with quark coalescence as the mechanism for hadronization. The paper is organized in the following way. In Section \textrm{2}, the coalescence mechanism and the AMPT model are briefly discussed. Section \textrm{3} describes our calculations for $D^{0}$ and $D^{\pm}_{S}$ $v_{2}$ at $\sqrt{s_{NN}}$ = 200 GeV and 2.76 TeV and a comparison with the measured results from LHC at 2.76 TeV. Finally, we summarize in Section \textrm{4}.

\section{The Quark Coalescence Model}

The AMPT model is a hybrid transport model~\cite{ampt,ampt1}. It uses the same initial conditions as in Heavy Ion Jet Interaction Generator (HIJING)~\cite{hijing}. Fragmentation in HIJING is based on the Lund string model. The Lund string fragmentation function is given by
\be
f(z) \propto z^{-1} (1-z)^{a} \exp(-b m_{T}^{2}/z),
\ee
where $z$ is the light-cone momentum fraction of the produced particle of transverse mass $m_{T}$ to that of the fragmenting string. The parameter $a$ and $b$  are used to fixed the measured charged particle multiplicity~\cite{lund}. The input parameters are available for RHIC and LHC energies based on previous studies. At the top RHIC energy, the value of $a$ = 2.2 and $b$ = 0.5 GeV$^{-2}$ (RHIC-tuned) is used to match the measured charged hadrons multiplicity with the value of QCD coupling constant ($\alpha_{s}$) being 0.47~\cite{rhic_chgv2}, whereas, for 2.76 TeV,  the value of $ a$ = 0.5 and $b$ = 0.9 GeV$^{-2}$ (LHC-tuned) is used to fit the data with $\alpha_{s}$ = 0.33~\cite{lhc_chgv2}. In this study, we have used RHIC-tuned and LHC-tuned values for the respective energies using version 2.26 of AMPT model.

In the string melting mode of AMPT model (labeled as AMPT-SM), strings are converted to soft partons. Scattering among partons are modeled by Zhang's parton cascade~\cite{ZPC}, which calculates two-body parton scatterings using cross section from pQCD with screening masses ($\mu$). The value of parton-parton scattering cross-section, $\sigma_{PP}$, is given by 
\be
\sigma_{PP} \approx \frac{9 \pi \alpha_s ^2}{2 \mu^2}.
\label{muaplha}
\ee
As hadronization of heavy quarks is not implemented in AMPT-SM, the phase-space information of partons at freezeout is taken from APMT-SM and used to form the open charm mesons according to the coalescence model described below.

Within the framework of the coalescence mechanism~\cite{coal}, the probability of producing a hadron from a soup of partons is determined by the overlap of the spatial and momentum space distribution of partons at equal time at freeze-out with the Wigner phase space function of partons inside the hadron under the assumption that the correlations between coalescing partons is weak and the binding energy of the formed hadron can be neglected. The coalescence mechanism used in this paper is described in great detail in ~\cite{cmco}.

The Wigner phase space function for quarks inside a meson is obtained from its constituent quark wave function
\begin{small}
\bea
\rho^W(\mathbf{r},\mathbf{k}) & = & \int \psi \left( \mathbf{r}+\frac{\mathbf{R}}{2} \right) \psi^{\star}\left( \mathbf{r}-\frac{\mathbf{R}}{2} \right) \exp(-i \mathbf{k} \cdot \mathbf{R}) d^3 \mathbf{R} \nonumber \\
 & = & 8 \exp(-\frac{r^2}{\sigma^2}-\sigma^2 k^2).
\eea
\end{small}
Here $R$ is the center-of-mass coordinate of the two quarks or antiquarks and $\psi^{\star}$ represents the complex conjugate of  quark wave function ($\psi$).
The relative momentum between the two quarks is $\mathbf{k} = (\mathbf{k}_1 - \mathbf{k}_2)/2$ and the quark wave function is given by spherical harmonic oscillator described as
\be
\psi (\mathbf{r}_1,\mathbf{r}_2) = \frac{1}{(\pi \sigma^2)^{3/4}} \exp \left[ \frac{-r^2}{2\sigma^2} \right],
\ee
with $\mathbf{r} = \mathbf{r}_1 - \mathbf{r}_2$ being the relative distance and $\sigma$ is the size parameter related to the root mean square radius as $\langle r^2 \rangle = (3/8)^{1/2} \sigma$. For the purpose of calculations in this paper, we have taken $\sigma = 0.47$ fm$^{2}$~\cite{cmco}. 
%%%%%%%%%%%%%%%%---Fig. 1----%%%%%%%%%%%%%%%%%%%%%%                                                                                       
\begin{figure}
\begin{center}
\includegraphics[scale=0.4]{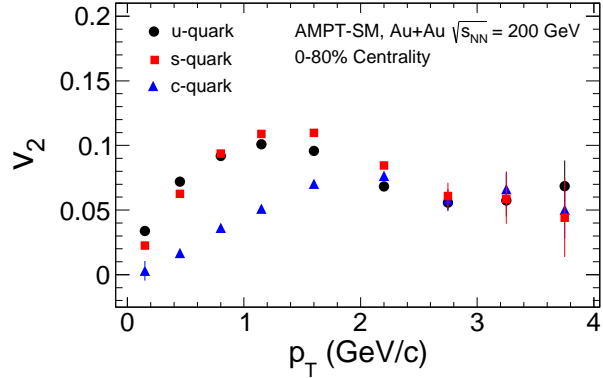}
\caption{Elliptic flow  of quarks as a function of
$p_{T}$  in minimum bias Au+Au collisions at
$\sqrt{s_{NN}}$ = 200 GeV using AMPT-SM model ($\sigma_{PP}$ = 10 mb).}
\label{quark_v2}
\end{center}
\end{figure}
%%%%%%%%%%%%%---------%%%%%%%%%%%%%%%%%%%%%%%%                                                                                                         

The $v_{2}(p_{T})$ of $u(d)$, $s$ and $c$ quarks which is  obtained using Eq.~\ref{2} from AMPT-SM model ($\sigma_{PP}$ = 10 mb) in minimum bias ($0-80\%$) Au+Au collisions at $\sqrt{s_{NN}}$ = 200 GeV and used as an input for the coalescence framework is shown in Fig.~\ref{quark_v2}. $u$ and $d$ quarks are assumed to be the same~\cite{nsm_ss}. We can see that $v_{2}$ of $u$ and $s$ quarks are comparable for all $p_{T}$, whereas $c$ quark has smaller $v_{2}$ for $p_{T}$ $<$ 2.0 GeV/c. This is consistent with the understanding that the charm quarks freeze early in the evolution of QGP due to their large mass. The $v_{2}(p_{T})$ of $u$, $s$ and $c$ quarks for AMPT-SM with $\sigma_{PP}$ = 3 mb is quantitatively smaller, but qualitatively similar (not shown here). 

%%%%%%%%%%%%%---------%%%%%%%%%%%%%%%%%%%%%%%%
\section{Results and Discussion}
\subsection{Elliptic flow of $D^0$ and $D_S ^\pm$ in Au+Au collisions at $\sqrt{s_{NN}}$ = 200 GeV}

%%%%%%%%%%%%%%%%---Fig.2&3----%%%%%%%%%%%%%%%%%%%%%%                                                                         
\begin{figure}
\begin{center}
\includegraphics[scale=0.4]{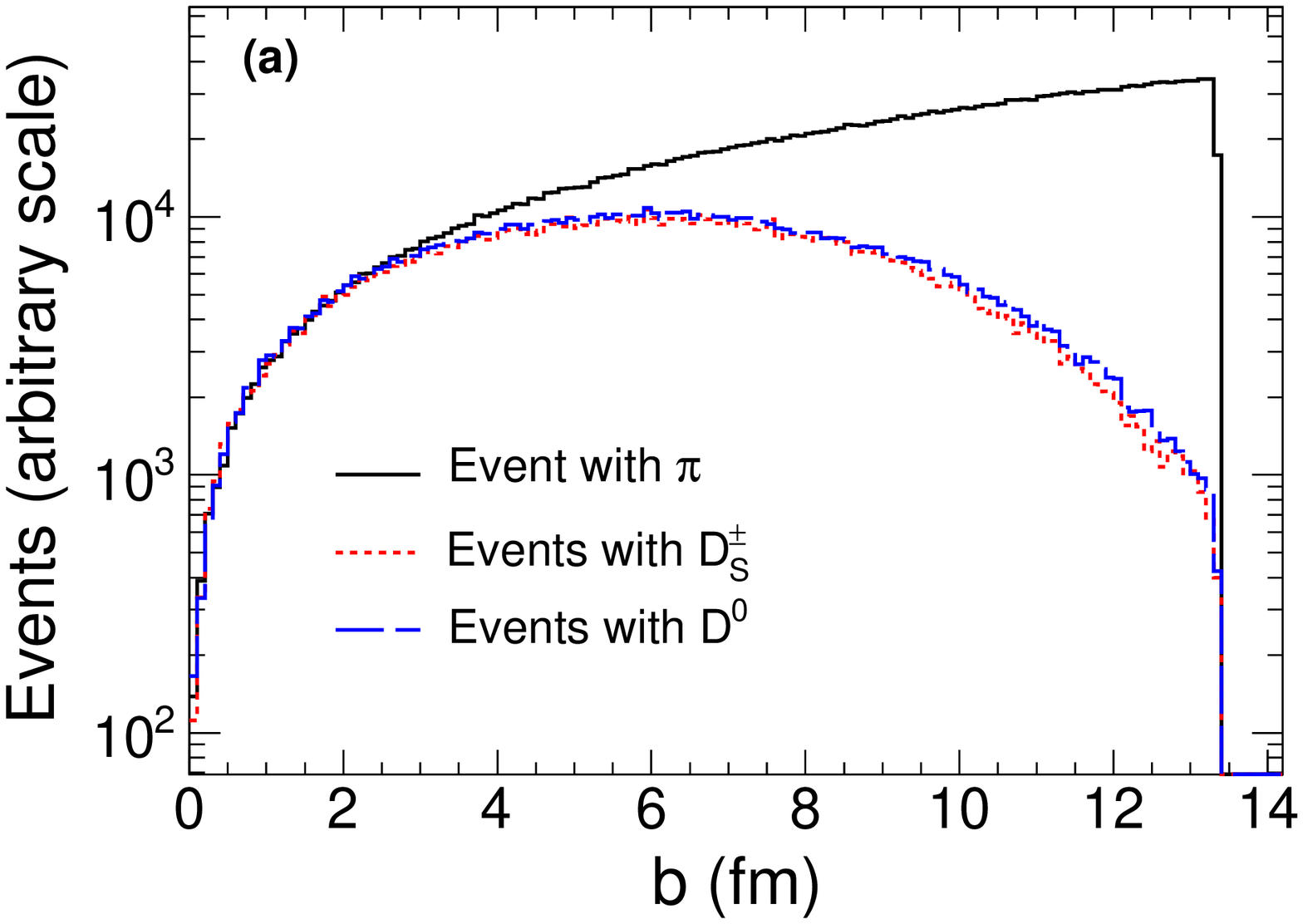}
\includegraphics[scale=0.4]{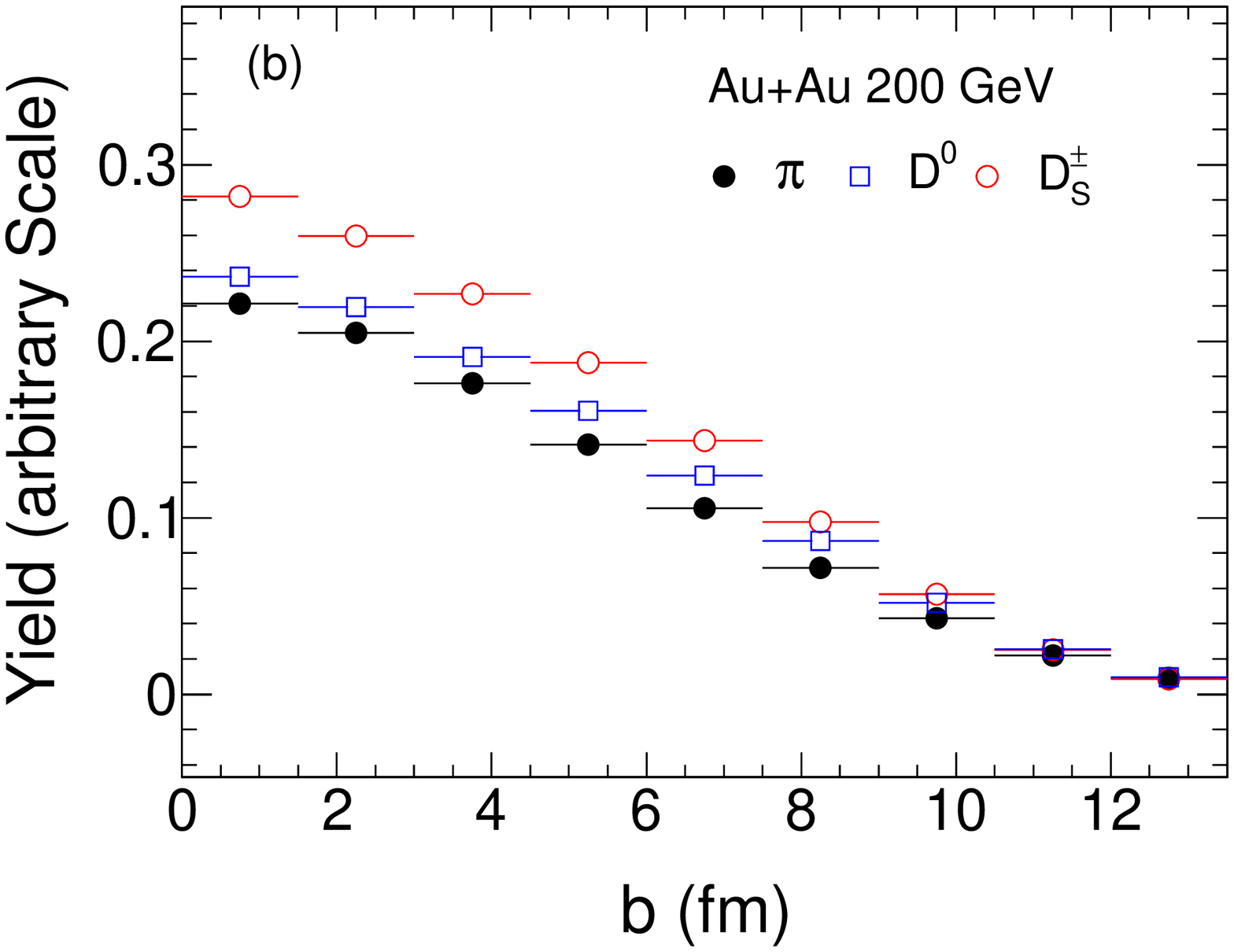}
\caption{Panel (a): Impact parameter ($b$) distribution for events with at least one $\pi$ (black), events with at least one $D^{0}$ (blue) and events with at least one $D^{\pm}_{S}$ (red) in Au+Au collisions at $\sqrt{s_{NN}}$ = 200 GeV. Y-axis is normalized for 0 $<$ $b$ $<$ 2 fm. Panel (b): Normalized yield of charged pion (black), $D^{0}$ (blue) and $D^{\pm}_{S}$ (red) as a function of $b$ in Au+Au collisions at $\sqrt{s_{NN}}$ = 200 GeV. Y-axis is normalized for 12 $<$ $b$ $<$ 13 fm.}
\label{event_yield}
\end{center}
\end{figure}
%%%%%%%%%%%%%%%%%%%%%%%%%%%%%%%%%%%%%%%%

%%%%%%%%%%%%%%%%%%%%%%%%%%%
\begin{figure}
\begin{center}
\includegraphics[scale=0.4]{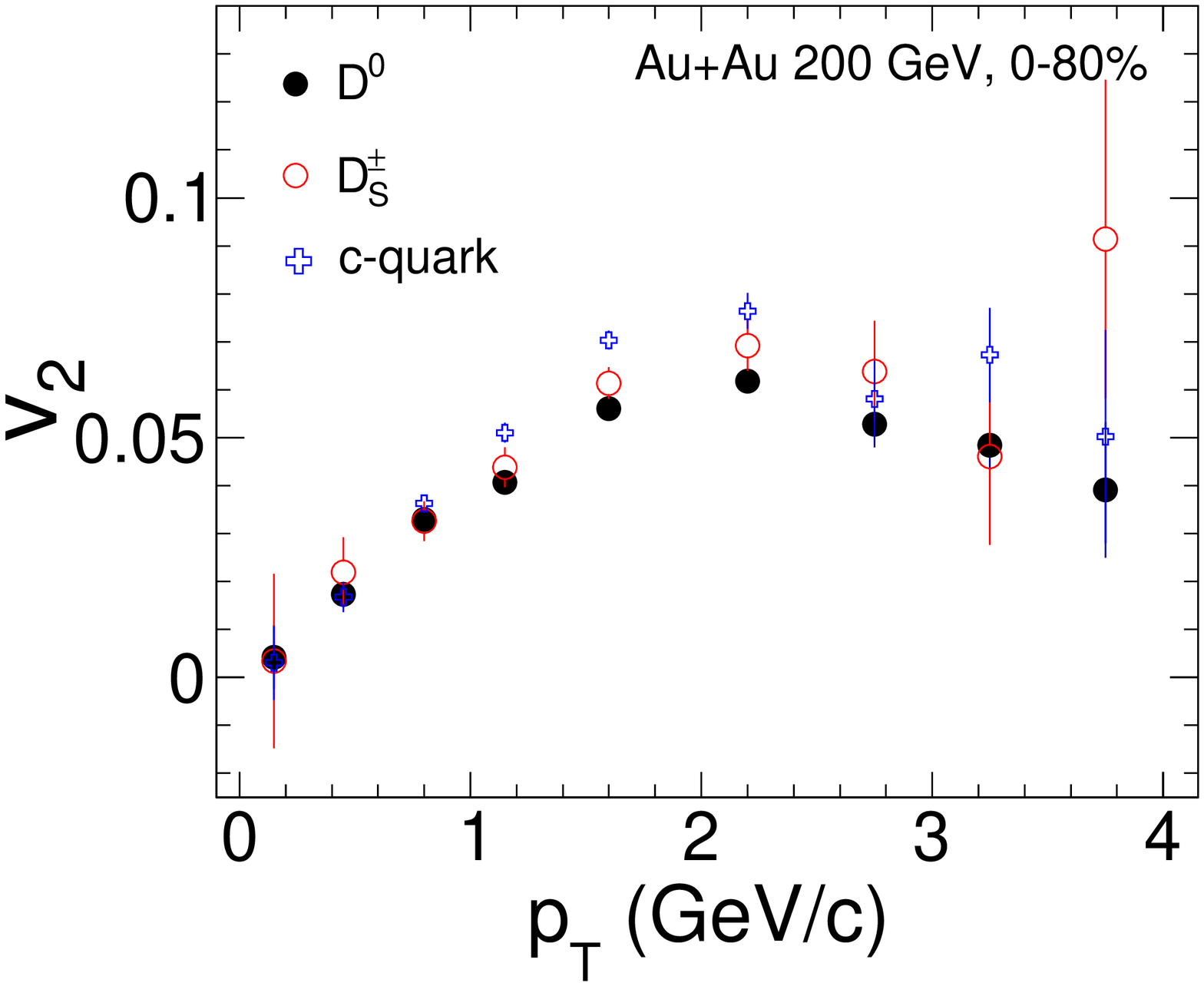}
\caption{Elliptic flow  of $D^{0}$, $D_S ^\pm$ and $c$ quarks as a function of $p_{T}$ in 0-80$\%$ minimum bias Au+Au collisions at $\sqrt{s_{NN}}$ = 200 GeV.}
\label{d0_ds_comp}
\end{center}
\end{figure}
%%%%%%%%%%%%%%%%---Fig.5----%%%%%%%%%%%%%%%%%%%%%% 

Fig.~\ref{event_yield} shows the distribution of number of events  and  particle yield for $D^0$, $D_S ^\pm$ and $\pi$ as function of impact parameter ($b$) in Au+Au collisions at $\sqrt{s_{NN}}$ = 200 GeV from coalescence model. In Fig.~\ref{event_yield}(a), the black, blue and red histograms correspond to the events which contain at least one $\pi$, $D^0$ and $D_S ^\pm$, respectively. The Y-axis has been normalized for impact parameter, $b$, between 0 to 2 fm for Fig.~\ref{event_yield}(a) and 12 to 13 fm for Fig.~\ref{event_yield}(b). The shape of the distribution of events with at least one open charm meson like $D^0$ and $D_S ^\pm$ is very different from those for pions and are biased towards central events. Similarly, the yield of $D^0$ and $D_S ^\pm$ is biased towards central collisions as shown in Fig.~\ref{event_yield}(b). This is because the charm quarks are produced in primary hard collisions and therefore the relative abundances of charm quarks over light quarks are more in central collisions than that of peripheral collisions. Since measured $v_{2}$ is an average over all particles in an event and over all events, a bias will be introduced if comparison is made between $v_{2}$ of open charm meson to that of charged hadron for a wide centrality bin, like 0-80$\%$. So, one should be very careful while comparing $v_{2}$ of open charm mesons ($D^0$, $D^\pm$, $D_S ^\pm$) with that of charged hadrons for a wide centrality range. However, we can compare $v_{2}$ open charm meson from experiment and model in a wide centrality range as the distribution of production probability of open charm meson as a function of centrality is roughly the same in our model. Having said that we still believe that measurements of $v_{2}$ should be done in smaller centrality classes. 
%%%%%%%%%%%%%---------%%%%%%%%%%%%%%%%%%%%%%%%                                          
\begin{figure*}
\begin{center}
\includegraphics[scale=0.8]{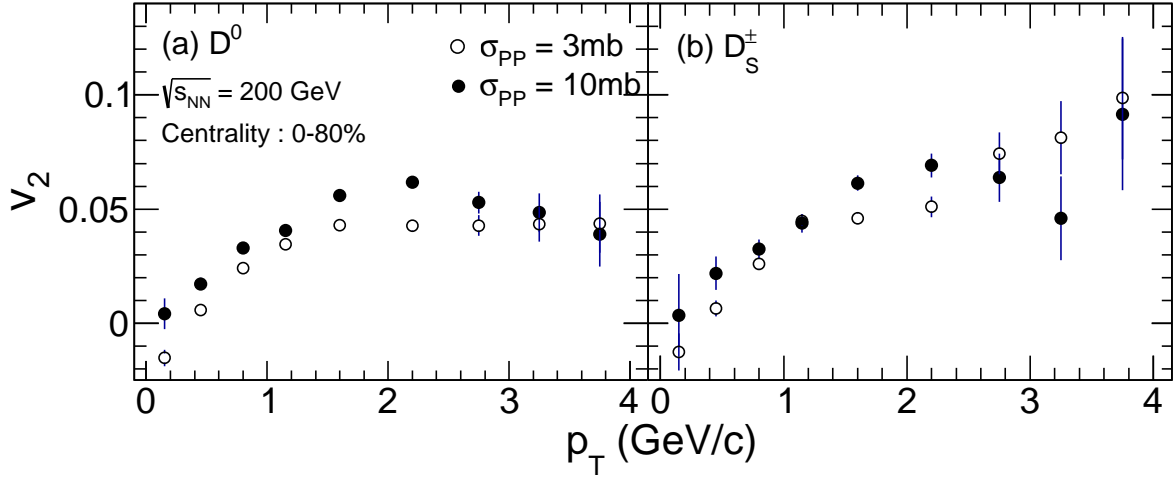}
\caption{Elliptic flow of $D^{0}$ and $D_S ^\pm$ as a function of $p_{T}$  in 0-80$\%$ minimum bias Au+Au collisions at $\sqrt{s_{NN}}$ = 200 GeV for $\sigma_{PP}$ = 3 mb and 10 mb.}
\label{charm_v2_08}
\end{center}
\end{figure*}
%%%%%%%%%%%%%---------%%%%%%%%%%%%%%%%%%%%%%%% 
The elliptic flow of $D^{0}$ and $D_S ^\pm$ mesons at mid-rapidity from the coalescence of quarks within the framework of AMPT-SM model is presented in Fig.~\ref{d0_ds_comp}. The results are for 0-80$\%$ minimum bias Au+Au collisions at $\sqrt{s_{NN}}$ = 200 GeV. The parton-parton interaction cross-section is 10 mb for this calculation. The centrality range was chosen to be consistent with the available preliminary data from the STAR experiment~\cite{star_d0}~\cite{star_ds}. 
 
The $v_{2}$ of charm quark is also shown in Fig.~\ref{d0_ds_comp}. The elliptic flow of both $D^{0}$ and $D_S ^\pm$ mesons follow the same trend as that of $c$ quark. In the AMPT model, the mass of $c$ quark is taken to be 1.35 GeV, which is much heavier than the light $u$ (6 MeV) and $s$ (199 MeV) quarks. In the coalescence mechanism, mass of quarks plays the role of the weight factor~\cite{mass_effect} and hence, the $v_{2}$ of $D^{0}$ and $D_S ^\pm$ meson is similar to $v_{2}$ of $c$ quark. The slight difference between the $v_{2}$ of $D^{0}$ and $D_S ^\pm$ meson at intermediate $p_{T}$ is due to the difference in $v_{2}$ of $u$ and $s$ quarks as shown in Fig.~\ref{quark_v2}. 

Results from RHIC  for $v_2$ of identified light hadrons as function of transverse momentum ($p_T$) shows that at intermediate $p_T$, the mesons and baryons form two different groups~\cite{ncq1}~\cite{ncq2}~\cite{ncq3}. When $v_2$ and $p_T$ are scaled by the number of constituent quarks of the hadron, the measured $v_2$ values are consistent with each other as predicted by the parton coalescence or recombination models. This observation is known as the number of constituents quark (NCQ) scaling. From Fig.~\ref{d0_ds_comp}, it should also be noted that NCQ scaling may not hold for $D^{0}$ and $D_S ^\pm$ as the elliptic flow of $c$ quark dominates the flow of the meson within the regime of this model.

The effect of parton-parton interaction cross-section on the $v_{2}$ of $D^{0}$ and $D_S ^\pm$ meson is shown in Fig.~\ref{charm_v2_08} where a comparison is made for minimum-bias Au+Au collisions at $\sqrt{s_{NN}}$ = 200 GeV with parton-parton interaction cross-section of 3 mb and 10 mb. There is a clear change in $v_{2}$ for $D^{0}$ and $D_S ^\pm$ and is quite similar to the previous observations for $v_{2}$ of charged hadrons from AMPT model~\cite{ampt1}. It has been observed that an interactions cross-section of 3--10 mb between partons explains charged hadrons $v_{2}$ in minimum bias Au+Au collisions at $\sqrt{s_{NN}}$ = 200 GeV. With this information, the comparison between data for open charmed $v_{2}$ and our calculated $v_{2}$ for 3 and 10 mb cross-section can be used to get more information about the parton-parton interaction cross-section of the system. In order to probe this, a high precision measurement of open charm $v_{2}$ will be needed.
               
\subsection{Elliptic flow of $D$ meson in Pb+Pb collisions at $\sqrt{s_{NN}}$ = 2.76 TeV}
\begin{figure}
\begin{center}
\includegraphics[scale=0.4]{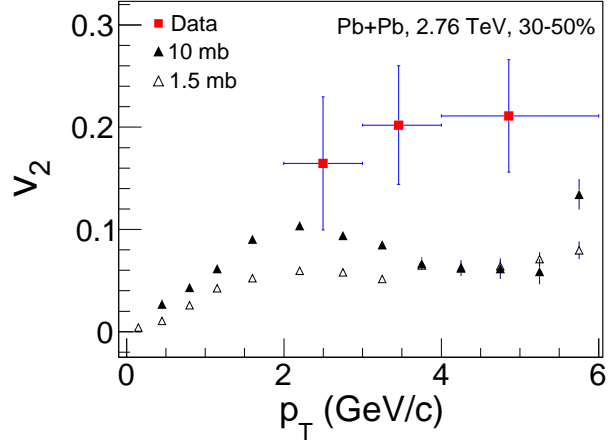}
\caption{Elliptic flow of  $D$ meson (average of $D^{0}$ and $D^{\pm}$) at mid-rapidity in Pb+Pb collision at  $\sqrt{s_{NN}}$ = 2.76 TeV for 30-50$\%$ centrality. Only statistical error is shown for ALICE data~\cite{alice_d0}.}
\label{lhcv2}
\end{center}
\end{figure}

Fig.~\ref{lhcv2} shows a comparison between our model calculations for $v_{2}$ of $D$ meson (average of $D^{0}$ and $D^{\pm}$) and measured $D$ meson $v_{2}$ at 2.76 TeV for 30-50$\%$ central collisions by the ALICE experiment~\cite{alice_d0}. In Fig.~\ref{lhcv2}, we have used LHC-tuned parameters for partonic cross-sections of 1.5 and 10 mb. Previous studies have shown that parton-parton interaction cross-section of 1.5 mb is sufficient to describe charged hadron $v_{2}$ for $p_{T} < 2$ GeV/c. We have seen that our model calculation for both 1.5 and 10 mb under-predict the data for $D$-meson $v_{2}$ for this tuning. This discrepancy at high $p_{T}$ may be related to the fact that particle production in the high $p_{T}$ region is dominated by parton fragmentation, while our model includes only recombination mechanism for particle production. It would be very interesting to see the behavior of data at low $p_{T}$ (below 2 GeV/c). Therefore, the results from future ALICE upgrade~\cite{alice_upgrade} will be very useful to study $v_{2}$ of both heavy flavor and charged hadrons at low $p_{T}$. \\
%%%%%%%%%%%%%%%%---Fig.5----%%%%%%%%%%%%%%%%%%%%%%                                                                         
%%%%%%%%%%%%%---------%%%%%%%%%%%%%%%%%%%%%%%%   
\subsection{Effect of shear viscosity on $v_{2}$}
%%%%%%%%%%%%%%%%---Fig.6----%%%%%%%%%%%%%%%%%%%%%%                                                                           
\begin{figure*}
\begin{center}
\includegraphics[scale=0.8]{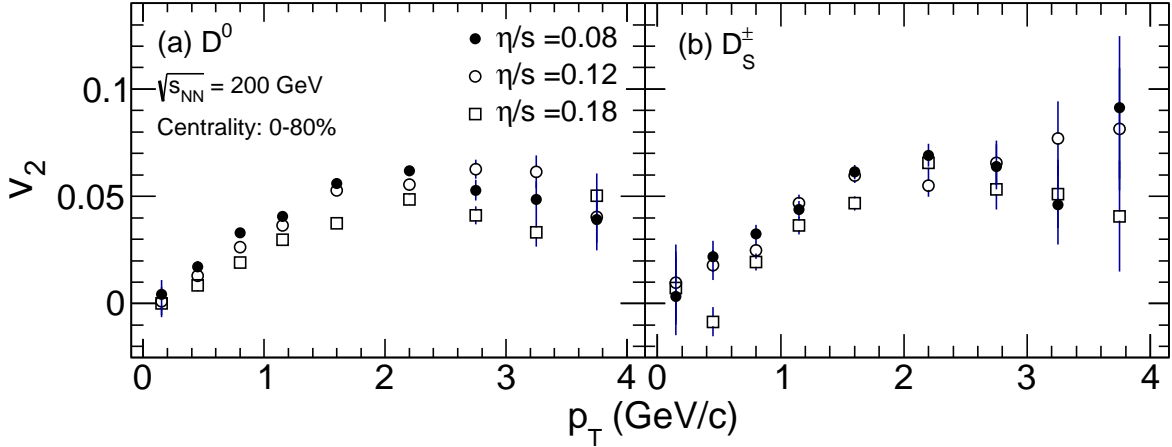}
\caption{Elliptic flow of $D^{0}$ and $D_S ^\pm$ as a function of $p_{T}$ for different viscous medium ($\eta_s/s$ = 0.08, 0.12 and 0.18) in minimum bias Au+Au collisions at $\sqrt{s_{NN}}$ = 200 GeV and $\sigma_{PP}$ = 10 mb.}
\label{charm_v2_08_visco}
\end{center}
\end{figure*}
%%%%%%%%%%%%%---------%%%%%%%%%%%%%%%%%%%%%%%% 

\begin{table}
\begin{center}
\caption{Values of $\eta_s/s$ for different values of $\alpha_s$ and $\mu$, keeping $\sigma_{PP}$ = 10mb for Au+Au collisions at $\sqrt{s_{NN}}$ = 200 GeV.}
\begin{tabular}{|c|c|c|}
\hline
  Shear viscosity to entropy density ratio, & QCD coupling constant, & Screening mass, \\
 $\eta_s/s$ & $\alpha_s$ & $\mu$ (in fm$^{-1}$) \\ \hline
  0.08  &  0.47  &  1.77  \\ \hline
  0.12  &  0.33  &  1.24  \\ \hline
  0.18  &  0.23  &  0.88  \\ \hline

\end{tabular}
\label{input_para}
\end{center}
\end{table}

Transport coefficients play a major role in probing the properties of the soup of quarks and gluons created in high energy heavy ion collisions~\cite{trans_1,trans_2,trans_3,trans_4,trans_5,trans_6,trans_7}. In order to study this, we compare the behavior of $v_2$ of open charm mesons for different values of the ratio of shear viscosity, $\eta_s$, to entropy density, $s$, with $p_{T}$ in Fig.~\ref{charm_v2_08_visco}. For a system of massless quarks and gluons at temperature  $T$ ($T = 378$ MeV at RHIC energy and $T = 468$ MeV at LHC energy in AMPT~\cite{lhc_chgv2}), the shear viscosity to entropy  density ratio is given by~\cite{lhc_chgv2}
\be
\frac{\eta_s}{s} \approx \frac{3\pi}{40 \alpha_s ^2} \frac{1}{\left( 9+ \frac{\mu^2}{T^2} \right) \ln \left( \frac{18 + \mu^2/T^2}{\mu^2/T^2} \right) - 18}. 
\label{eq_visco}
\ee
In the AMPT model, the magnitude of $\frac{\eta_s}{s}$ can be done by tuning $\alpha_s$ and $\mu$, keeping $\sigma_{PP}$ fixed according to Eq.~\ref{muaplha} and~\ref{eq_visco}.
Fig.~\ref{charm_v2_08_visco} shows the change in $v_{2}$ of $D^{0}$ and $D_S ^\pm$ meson due to the variation of $\eta_s/s$ in Au+Au collisions at $\sqrt{s_{NN}}$ = 200 GeV. We have used three different value of $\eta_s/s$: 0.08, 0.12, and 0.18. Table~\ref{input_para} gives the values of $\alpha_s$ and $\mu$ for different values of $\eta_s/s$. 
As Eq.~\ref{eq_visco} is approximate, the values of $\eta_s/s$ obtained in Fig.~\ref{charm_v2_08_visco} may differ slightly from the actual values from data. In our case, we wanted to illustrate the change in $v_{2}$ due to variation in $\eta_s/s$ and our conclusions are, therefore, independent of the approximations.
In Fig.~\ref{charm_v2_08_visco}, we show that $v_2$ decreases with increase in $\eta_s/s$. This is consistent with the interpretation that increased shear viscosity reduces anisotropic expansion and hence reduces $v_2$. The production and propagation of open charm meson in QGP medium is expected to be different from light hadrons because of their large mass. Therefore, a study of the effect of $\eta_s/s$ on $v_{2}$ for both open charm mesons and charged hadrons will be a useful to probe of the QGP medium. In addition, this can give us a constraint on the $\eta_s/s$ of the relatively early time scale of the medium created in relativistic heavy ion collisions.
   
\begin{figure}
\begin{center}
\includegraphics[scale=0.4]{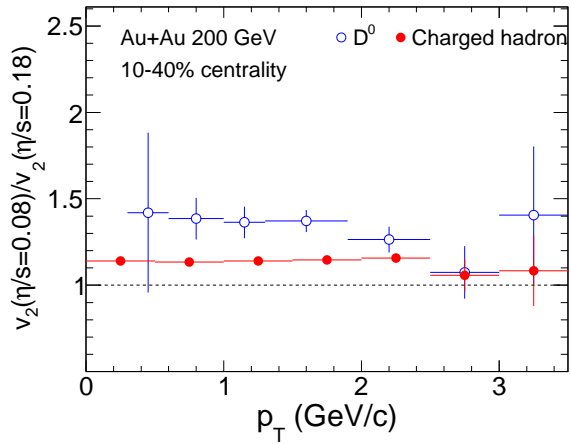}
\caption{ Ratio between $v_{2}$ for $\eta_s/s$ = 0.08  and for $\eta_s/s$ = 0.18  as a function of $p_{T}$ in Au+Au collisions at $\sqrt{s_{NN}}$ = 200 GeV for charged hadrons and $D^{0}$ for 10--40\% central collisions.}
\label{ratio_visco}
\end{center}
\end{figure}

The ratio of $v_{2}$ for $\eta_s/s$ = 0.08 and for $\eta_s/s$ = 0.18 is shown as function of $p_{T}$ for 10--40\% central collisions in Fig.~\ref{ratio_visco}. The solid red and open blue circles represents the results for charged hadrons and $D^{0}$ respectively. We can see that the change in $v_{2}$ for charged hadrons is $\sim$15$\%$, whereas for $D^{0}$, it lies between 30--40$\%$ for $p_{T}$ $<$ 2.0 GeV/c. We find this ratio to be independent of $p_{T}$, centrality and energy in our model. We have also observed the results for $D_S ^\pm$ (not shown) to be similar to $D^{0}$. Hence, we conclude that the elliptic flow of open charm meson is more sensitive to viscous properties of the QGP medium compared to the light charged hadrons. 
We have also verified the conclusion of Fig.~\ref{ratio_visco} in a different approach, where we  varied the values of $\eta_s/s$  by keeping $\alpha_s$ fixed but changing $\mu$.
High precision measurement of open charm $v_{2}$, mainly at low $p_{T}$, is needed at RHIC and LHC to probe the transport properties of the medium created in heavy-ion collision.
%%%%%%%%%%%%%%%%---Fig.8----%%%%%%%%%%%%%%%%%%%%%%                                                                         
%%%%%%%%%%%%%%%%-------%%%%%%%%%%%%%%%%%%%%%%                                                                           
\section{Summary and Conclusion}
We have studied the elliptic flow of open charm mesons as a function of $p_T$ for heavy ion collisions at $\sqrt{s_{NN}}$ = 200 GeV and 2.76 TeV with A Multi Phase Transport model. We have implemented quark coalescence as the mechanism for hadron production for open charm mesons in this model study. We have given predictions for $v_{2}(p_{T})$ of $D^{0}$ and $D_S ^\pm$ meson for minimum-bias Au+Au collision at RHIC for different values of partonic interaction cross-sections. We find that $v_2$ increases with increase in the partonic interaction cross-section. We have shown that the production of open charm mesons is heavily biased towards central events as compared to light hadrons, therefore, one should use appropriate weights when comparing $v_{2}$ of open charm mesons with light hadrons for wide centrality, like 0-80$\%$. 

We have also shown that parameters tuned to describe charged hadron multiplicity and $v_{2}$ at 2.76 TeV fail to reproduced $D$ meson $v_{2}$ in Pb+Pb collisions at $\sqrt{s_{NN}}$ = 2.76 TeV for $p_{T}$ $>$ 2 GeV/c. Precision measurement of $D$-meson $v_{2}$ is needed at low $p_{T}$ to extract more quantitative properties of the medium formed at LHC.
 
We have also presented a systematic study on the effect of $\eta_s/s$ on elliptic flow within the transport model approach. There seems to be a considerable decrease in $v_2$ with an increase in $\eta_s/s$.  We have also shown that open charmed meson are more sensitive to viscous properties of QGP medium compared to light charged hadrons.

%\section*{Acknowledgement}
\ack{
We thank Zi-Wei Lin for useful discussion on the AMPT model. Financial support from US DOE Office of Science is gratefully acknowledged.\\
}
%\begin{thebibliography}{99}

\end{document}